\documentclass[preprint]{ptptex}

\usepackage[dvips]{graphicx}
\usepackage{amsmath,amsthm,amssymb}
\usepackage{amsbsy}

\newcommand{\bZ}{\mathbb{Z}}

\def\e{{\rm e}}
\def\m{{\rm m}}
\def\half{\frac12}

\def\mbf{\boldsymbol}

\def\nn{\nonumber\\}
\def\Im{{\rm Im}}
\def\Re{{\rm Re}}
\def\D{{\rm D}}
\def\o{{\rm o}}
\def\bE{{\mbf E}}
\def\bA{{\mbf A}}
\def\bB{{\mbf B}}

\def\bDdp{{\mbf D}_{\rm dipole}}
\def\bD{{\mbf D}}
\def\bS{{\mbf S}}
\def\bsigma{{\mbf \sigma}}
\def\CASE#1#2{\begin{cases}\displaystyle#1\\ \displaystyle#2\end{cases}}
\def\mypmatrix#1{\begin{pmatrix}#1\end{pmatrix}}
\def\half{{1\over2}}
\def\hatr{{\hat r}}
\newcommand{\abs}[1]{\left|{#1}\right|}

\notypesetlogo                       
\preprintnumber[3cm]{
KEK-TH-1166\\
YITP-07-48\\ August, 2007}

\title{
Electric Dipole Moments of Dyon and `Electron'%
}

\author{
Makoto \textsc{Kobayashi},$^{1,2,}$\footnote{%
E-mail:\  {\tt kobayath@post.kek.jp}}
\ Taichiro \textsc{Kugo}$^{3,}$\footnote{%
E-mail:\  {\tt kugo@yukawa.kyoto-u.ac.jp}}
and 
Tatsuya \textsc{Tokunaga}$^{3,}$\footnote{%
E-mail:\  {\tt tokunaga@yukawa.kyoto-u.ac.jp}}
}

\inst{
$^{1}$High Energy Accelerator Research Organization (KEK), \\
Tsukuba 305-0801, Japan \\
$^2$
International Institute for Advanced Studies, Kizugawa 619-0225, Japan \\
$^{3}$
Yukawa Institute for Theoretical Physics, Kyoto University, \\
Kyoto 606-8502, Japan
}

\abst{
The electric and magnetic dipole moments of dyon fermions are calculated 
within $N=2$ supersymmetric Yang-Mills theory including the $\theta$-term. 
It is found, in particular, that the gyroelectric ratio deviates from 
the canonical value of 2 for the monopole fermion $(n_\m{=}1,n_\e{=}0)$ 
in the case $\theta\not=0$. Then, applying the $S$-duality transformation 
to the result for the dyon fermions, we obtain 
an explicit prediction for the electric dipole moment (EDM) of the 
charged fermion (`electron'). It is thus seen that the approach 
presented here provides a novel method for computing the EDM induced by 
the $\theta$-term. }

\begin{document}

\maketitle

\section{Introduction}

The $\theta$-term in gauge theory violates the CP symmetry, and hence it 
can generally induce an electric dipole moment (EDM) for a 
particle with spin.
However, at present, there exists no general, established method to 
calculate this EDM. The reason for the difficulty is that 
the $\theta$-term is a total derivative and therefore it neither changes 
the equation of motion nor contributes to the Feynman rules. Thus no 
calculation in which the $\theta$-term itself is treated perturbatively 
can provide a method for determining the EDM. In the conventional 
calculation of the EDM of the neutron in QCD, this problem is avoided 
by first performing a 
chiral transformation to replace the $\theta$-term with the 
complex phase of the quark mass term through the chiral $U(1)$ anomaly. 
Then, because such a mass term with a complex phase is readily treated 
in a perturbative calculation, the EDM can be computed. In this 
computation, the anomaly gives a non-perturbative information. 

Here we propose an interesting and novel method to calculate 
the EDM of a charged fermion.    
We consider $N=2$ $SU(2)$ supersymmetric Yang-Mills theory, in which 
there exists a BPS monopole multiplet consisting of a 
monopole scalar and a monopole fermion ($J=1/2$). 
We apply the $S$-duality transformation to this system, and as a result, 
the monopole multiplet turns to a charged BPS multiplet
consisting of a scalar and a spin 1/2 fermion, which we call an 
`electron', for brevity. Note that the two systems before and after the 
$S$-duality transformation are distinct because their vacua 
differ. To distinguish these systems, we refer to them as the 
`monopole world' and the `electron world', respectively.

We compute the EDM of the fermion as follows. First, in the original 
(i.e. monopole) world, we calculate the electromagnetic fields around a 
monopole fermion placed at the origin. When the $\theta$-term is added, 
the monopole acquires an electric charge (through 
Witten effect\cite{Witten:1979ey}), in addition 
to the magnetic charge, and thus it becomes a dyon. We use the Julia-Zee 
classical dyon solution corresponding to its electric and magnetic 
charges. The electromagnetic fields around the monopole fermion can be 
found by applying the supersymmetry (SUSY) transformation to the 
classical Julia-Zee dyon solution. As discussed in 
previous works,\cite{Kastor:1998ma,Kobayashi:2006tv} the first-order term 
in the SUSY transformation parameter gives a fermion zero-mode solution 
representing the fermion partner of the monopole. The second-order term 
gives the desired electromagnetic field around the monopole 
fermion. We can determine the magnetic dipole moment and the electric 
dipole moment carried by the monopole fermion directly from the asymptotic 
form of this electromagnetic field.

Next we apply the $S$-duality transformation to the monopole system. The
BPS monopole multiplet is thereby transformed to the BPS electron 
multiplet in
the electron world. We then obtain the electromagnetic field around the 
electron by applying the $S$-duality transformation to that 
around the monopole fermion. The EDM, $\mu_\e$, and 
magnetic (dipole) moment, $\mu_\m$, can be determined from the asymptotic 
form of this electromagnetic field. Our result is the following: 
\begin{equation}
\mu_\m = {e\over m}J, \qquad \mu_\e = {e^2\theta\over8\pi^2}{e\over m}J,
\end{equation}
with $J=1/2$ standing for the spin. 
The magnetic (dipole) moment $\mu_\m=2\times(e/2m)J$ is identically the 
Dirac value, corresponding to the gyromagnetic ratio $g_\m=2$. 
By contrast, 
the EDM is given by the `Dirac value', $eJ/m$, multiplied by a factor 
of $e^2\theta/8\pi^2$.

This paper is organized as follows. In \S 2, we compute the 
electromagnetic fields around a dyon fermion by performing a finite 
supersymmetry transformation of the Julia-Zee dyon solution, and we thus 
obtain both the EDM and the magnetic moment of the dyon fermion. 
In \S3, we elucidate the $SL(2;\bZ)$ transformation of the 
electromagnetic field strength. Then, we carry out the $S$-duality 
transformation, converting the monopole fermion state into an electron 
state, and find the EDM of the electron. The final section is devoted 
to discussion of the results.

\section{Electric dipole moment of a dyon}

We consider $N=2$ $SU(2)$ supersymmetric Yang-Mills theory with  
the Lagrangian
\begin{eqnarray}
{\cal L} &=& {\rm Tr} \left( - \frac{1}{4} F_{\mu\nu} F^{\mu\nu} 
+ \frac{1}{2} D^{\mu} S D_{\mu} S + \frac{1}{2} D^{\mu} P D_{\mu} P + \frac{e^2}{2} [S,P]^2 \right. \nonumber \\
  & &   ~~~~ \left.  + i \bar{\psi} \gamma^{\mu} D_{\mu} \psi- e \bar{\psi} [S,\psi] + i e \bar{\psi} \gamma_5 [P , \psi] 
+ \frac{\theta e^2 }{32 \pi^2} F_{\mu\nu} {\tilde{F}}^{\mu\nu}  \right) ,
\label{eq:L}
\end{eqnarray}
where all fields are written in matrix form; i.e., here we have 
$S=S^aT^a$, etc., with the 
generators $T^a$ in the fundamental representation satisfying the relation 
${\rm tr}\,T^aT^b = (1/2)\delta^{ab}$. We define the operator Tr as 
${\rm Tr}\equiv2\,{\rm tr}$, so that ${\rm Tr}\,T^aT^b = \delta^{ab}$. 
This Lagrangian is invariant, 
up to total derivative terms, under the $N=2$ supersymmetry 
transformations
\begin{eqnarray}
\delta A_{\mu} &=& i \bar{ \tilde{\alpha}} \gamma_{\mu} \psi- i \bar{\psi} \gamma_{\mu} \tilde{\alpha} , \nn
\delta S\phantom{\mu} &=& i \bar{ \tilde{\alpha}} \psi- i \bar{\psi} \tilde{\alpha} , \nonumber \\
\delta P\phantom{_\mu} &=& \bar{ \tilde{\alpha}}  \gamma_5 \psi- \bar{\psi} \gamma_5  \tilde{\alpha} , \nonumber \\
\delta\psi\phantom{\mu}&=& \left( \half \gamma^{\mu\nu} F_{\mu\nu} - \gamma^{\mu} D_{\mu} S 
+ i  \gamma^{\mu} D_{\mu} P \gamma_5 - e [P, S] \gamma_5 \right) \tilde{\alpha} ,  
\label{trans} 
\end{eqnarray}
where $\gamma^{\mu\nu}= \gamma^{[\mu}\gamma^{\nu]}\equiv(1/2)[\gamma^\mu,\gamma^\nu]$, and 
$\tilde{\alpha}$ is a Grassmann-valued Dirac spinor.\footnote{
We employ the convention in which $\gamma^\mu\gamma^\nu+\gamma^\nu\gamma^\mu=2\eta^{\mu\nu}, \ \eta^{\mu\nu}
=\hbox{diag}(+1,-1,-1,-1)$, and $\gamma^5=i\gamma^0\gamma^1\gamma^2\gamma^3$.}  
Witten showed in Ref.~\citen{Witten:1979ey} that the $\theta$-term 
induces an electric charge of $+e\theta/ 2 \pi$ for a magnetic monopole 
with unit magnetic charge, $g=4\pi/e$.
\footnote{
We have chosen the sign of the $\theta$-term in the Lagrangian (\ref{eq:L}) 
such that the Witten charge is $+\theta e / 2 \pi$.  This is the same as 
the convention of Alvarez-Gaum\'e and Hassan\cite{AlvarezGaume:1996mv}, 
but it is opposite those of Seiberg and Witten\cite{Seiberg:1994rs} 
and Harvey.\cite{Harvey:1996ur}
}
thereby implying the existence of dyons. 
In accordance with this fact, a classical dyon solution exists that  
has long been known as the Julia-Zee dyon\cite{Julia:1975ff}: 
\begin{eqnarray}
A^a_{ 0} &=& \frac{{\hat{r}}^a}{er} H(evr)\sinh\gamma, \qquad 
A^a_{i} = {\epsilon}^a {}_{i j} {\hat{r}}^j \frac{ 1 - K(e v r) }{e r}  , \nn
S^a &=& \frac{{\hat{r}}^a}{er} H(evr)\cosh\gamma, \qquad  \psi^a = P^a = 0,
\label{eq:JZsolution}
\end{eqnarray}
where the functions $K(x)$ and $H(x)$ are given by 
\begin{eqnarray}
K(x) &=& \frac{x}{\sinh x} ,\\
H(x) &=&  x \coth x - 1.
\label{eq:KHproperty}
\end{eqnarray}

The electric and magnetic charges, $Q_\e$ and $Q_\m$,  
of the unbroken $U(1)$ gauge field $F_{\mu\nu}
\equiv S^a\cdot F^a_{\mu\nu}/a$, with 
$a\equiv\sqrt{S^aS^a}$ at the spatial infinity, are defined by
\begin{eqnarray}
Q_\e &\equiv& \int_{S_\infty} \bE\cdot d\bS = {1\over a}\int d^3x\, \bE^a\cdot (\bD S)^a, 
\nn
Q_\m &\equiv& \int_{S_\infty} \bB\cdot d\bS = {1\over a}\int d^3x\, \bB^a\cdot (\bD S)^a. 
\label{eq:QeQm}
\end{eqnarray}
Here we have defined the quantities
\begin{equation}
\bE^a \equiv (F_{01}^a,\,F_{02}^a,\,F_{03}^a),\qquad 
\bB^a \equiv (-F_{23}^a,\,-F_{31}^a,\,-F_{12}^a).
\end{equation}
Inserting the solution (\ref{eq:JZsolution}) into the definition of $a$ 
and Eq.~(\ref{eq:QeQm}), we find that $a=v\cosh\gamma$ 
and the Julia-Zee dyon carries 
\begin{equation}
Q_\m=\frac{4 \pi}{e}\equiv g, \qquad  Q_\e= -\frac{ 4 \pi}{e }\sinh\gamma.
\label{eq:dyoncharge}
\end{equation}
Here, $g$ is the unit magnetic charge of monopole. Classically, the 
parameter $\gamma$ is an arbitrary real constant. However, as argued by 
Witten, the generator $N$ of the unbroken $U(1)$ gauge transformation $
\delta A_\mu^a = D_\mu S^a/ea$, which is found using the Noether method to be 
\begin{equation}
N= {\partial{\cal L}\over\partial(\partial_0A^a_\mu)}\delta A_\mu^a =
\frac{Q_\e}{e} -  \frac{ \theta e Q_\m }{8 \pi^2},  \label{U(1)generator}
\end{equation}
should have integer eigenvalues $n_\e$.   
Using the relation $eQ_\m=4\pi$ in (\ref{eq:dyoncharge}) for the 
magnetic charge of the Julia-Zee dyon, we see that the parameter $\gamma$, 
or equivalently, the electric charge $Q_\e$, is quantized 
as follows in the case that the Lagrangian includes 
the CP-violating $\theta$-term:
\begin{equation}
Q_\e= -{4\pi\over e}\sinh\gamma= n_\e e + \frac{ e\theta}{2 \pi}. \label{relation}
\end{equation}

The dyon mass is given by 
\begin{eqnarray}
m &=& \int d^3x \,\half \Bigl[ (\bE^a)^2 + (\bB^a)^2 + (\bD S^a)^2 
+ (D_0 S^a)^2\Bigr] \nn
&=& \int d^3x \,
\half \Bigl[ (\bE^a{-}\bD S^a\sin\alpha)^2\! 
+ (\bB^a{-}\bD S^a\cos\alpha)^2\! + (D_0 S^a)^2\Bigr] \nn
&& \hspace{4eM}{}+ a (Q_\e\sin\alpha+ Q_\m\cos\alpha),
\label{eq:Mass}
\end{eqnarray}
and thus it satisfies the relation $m\geq a(Q_\e\sin\alpha+ Q_\m\cos\alpha)$ 
for any $\alpha$. Because the quantity on the right-hand side is maximal 
when $\tan\alpha=Q_\m/Q_\e$, we obtain a bound, called 
Bogomol'nyi bound, expressed by the following: 
\begin{equation}
m \geq a \sqrt{Q_\m^2 + Q_\e^2}.
\end{equation}
The Julia-Zee dyon is characterized as the BPS state for which 
the  bound is saturated, 
and hence it satisfies the first-order Bogomol'nyi equations
\begin{equation}
\bB^a=\bD S^a\cos\alpha, \qquad  \bE^a=\bD S^a\sin\alpha, \qquad D_0 S^a =0,
\label{eq:Bogomol'nyi}
\end{equation}
with $\cos\alpha=Q_\m/\sqrt{Q_\m^2 + Q_\e^2}$ and 
$\sin\alpha=Q_\e/\sqrt{Q_\m^2 + Q_\e^2}$. Substituting the expressions 
(\ref{eq:dyoncharge}) for $Q_\e$ and $Q_\m$, we find the dyon mass and the relation between 
angles $\alpha$ and $\gamma$:
\begin{equation}
m_{\rm dyon}= a{4\pi c\over e}= gvc^2, \qquad 
\cos\alpha= {1\over c},\qquad \sin\alpha=-{s\over c}, 
\label{eq:dyonmass}
\end{equation}
where we have introduced abbreviations $c\equiv\cosh\gamma$ and $s\equiv\sinh\gamma$.
The Bogomol'nyi equations (\ref{eq:Bogomol'nyi}) 
are indeed 
satisfied by the following properties of functions $K$ and $H$:
\begin{equation}
x K' =  - KH , \qquad x H' = H - ( K^2 - 1). 
\end{equation}

It should be noted that the BPS state considered here preserves half of 
the supersymmetry, as is generally the case for BPS solutions. 
To show this explicitly, we use the following explicit representation 
for the $\gamma$-matrices:
\begin{equation}
\gamma^0 = 
\begin{pmatrix}
0 & -i \\
i & 0
\end{pmatrix}
=1\otimes \sigma_2,\qquad 
{\mbf\gamma}=  
\begin{pmatrix}
-i{\mbf\sigma} & 0 \\
0 & i{\mbf\sigma}
\end{pmatrix}
={\mbf\sigma}\otimes (-i\sigma_3).
\end{equation}
Then, using the dyon solution (\ref{eq:JZsolution}) and the 
relations 
\begin{equation}
\bD S^a= c \bB^a, \qquad  \bE^a= -s\bB^a
\label{eq:SB-EBrel}
\end{equation}
which follow from the Bogomol'nyi equations (\ref{eq:Bogomol'nyi}), 
the supersymmetry transformation 
$\delta\psi$ of the fermion in (\ref{trans}) 
takes the form
\begin{equation}
\delta\psi^a = 
\begin{pmatrix}
i(\bB^a +\bD S^a)\cdot \bsigma & \bE^a\cdot\bsigma \\
\bE^a\cdot\bsigma & 
i(\bB^a -\bD S^a)\cdot \bsigma 
\end{pmatrix}
\tilde\alpha 
=2i\bB^a\cdot\bsigma \otimes P_+(\gamma) \tilde\alpha,
\label{eq:2.18}
\end{equation}
where $P_+(\gamma)$ is a projection operator. It and the other projection 
operator, $P_-(\gamma)$, are given by  
\begin{equation}
P_\pm(\gamma) \equiv\half 
\begin{pmatrix}
1\pm c & \pm is \\
\pm is & 1 \mp c  
\end{pmatrix}
= {1\pm(c\sigma_3 +is\sigma_1)\over2}
=e^{{\gamma\over2}\sigma_2}\left({1\pm\sigma_3\over2}\right)e^{-{\gamma\over2}\sigma_2}.
\end{equation}
From Eq.~(\ref{eq:2.18}), it is seen that only half of the supersymmetry 
parameter, constituted by $P_+(\gamma)\tilde\alpha$,  is involved in $\delta\psi$, 
while the remaining half, $P_-(\gamma)\tilde\alpha$, corresponds to the 
unbroken supersymmetry.

Now, let us perform a finite supersymmetry transformation of the 
Julia-Zee dyon solution with the following form of the transformation 
parameter $\tilde\alpha$, parametrized by a two-component Grassmann 
parameter $\alpha$:
\begin{equation}
\tilde \alpha\equiv 
P_+(\gamma) 
\begin{pmatrix}
\alpha\\ 0 
\end{pmatrix}
=
\half \,\alpha\otimes 
\begin{pmatrix}
1+c \\ is 
\end{pmatrix}.
\end{equation} 
Any field $\Phi$ is transformed under  a finite supersymmetry 
transformation as 
\begin{eqnarray}
\tilde\Phi&=& e^{i({\bar{\tilde\alpha}}Q+\bar Q\tilde\alpha)}\Phi 
e^{-i({\bar{\tilde\alpha}}Q+\bar Q\tilde\alpha)}  \nn
&=& 
\Phi+ \delta\Phi+ \half \delta^2\Phi+ {1\over3!}\delta^3\Phi+ {1\over4!}\delta^4\Phi.
\end{eqnarray}
This series terminates at the fourth-order term in $\alpha$ since it is 
a two-component complex Grassmann parameter. 
In the computations involved in the present transformation of the 
classical solution, it is easier to calculate 
$\delta^n\Phi$ in the form $\delta^{n-1}(\delta\Phi)$ than  $\delta(\delta^{n-1}\Phi)$, because  
we can substitute the classical solution only at the end of the 
calculation, therefore we can use the result of the preceding step, 
$\delta^{n-1}\Phi$, only in the final step. 

Any solution of the equation of motion is transformed to another solution 
under a finite symmetry transformation, since the action is invariant under such a transformation. Therefore, the image of the Julia-Zee dyon 
solution under a finite supersymmetry transformation is also an exact solution of the equation of motion for any choice of 
the transformation parameter $\alpha$. The first-order term 
in $\alpha$ of $\tilde\Phi(\alpha)$ gives the massless fermion solution around the 
Julia-Zee dyon background. The higher-order terms describe the 
back-reaction of the created fermion.
The computation of these terms is carried out in Refs.~\citen{Kastor:1998ma} and \citen{Kobayashi:2006tv} for the case of the monopole solution 
(i.e., the $\theta=0$ case). In those works, it is found that the 
gyroelectric ratio, $g_\e$, is just equal to the Dirac value, 2, for 
the monopole fermion. 

We now perform the same calculation for the Julia-Zee dyon case and, 
in particular, examine the electric dipole moment of the Julia-Zee dyon 
fermion to determine whether $g_\e=2$ holds also in the $\theta\not=0$ case. 

At first-order in $\alpha$, only the fermion, $\psi$, is nonvanishing; all 
the boson fields are zero because  the fermion field $\psi$ vanishes 
at zeroth order:
\begin{equation}
\delta^1\psi^a 
=i\bB^a\cdot\bsigma \alpha 
\otimes 
\begin{pmatrix}
1+c \\ is 
\end{pmatrix}, \qquad 
\delta^1A_\mu^a =\delta^1S^a=\delta^1P^a=0.
\label{eq:FirstOrder}
\end{equation}
Conversely, the fermion vanishes at second order, 
since all the bosons vanish at the first order:
\begin{equation}
\begin{pmatrix}
\delta^2A_0^a \\
\delta^2S^a \\
\delta^2P^a 
\end{pmatrix}
=2(1+c)(\alpha^\dagger\bB^a\cdot\bsigma\alpha)\times 
\begin{pmatrix}
-c \\ -s \\ 1 
\end{pmatrix}, \qquad 
\delta^2\bA^a =0, \quad 
\delta^2\psi^a =0. 
\label{eq:SecondOrder}
\end{equation}
Note that $\delta^2\bA^a$ vanishes because we are restricting the 
transformation parameter to that part projected by $P_+(\gamma)$. 
Then, at third order, only the fermion could be non-vanishing, but 
with a straightforward calculation using the relations among 
$\delta^2A_0^a,\  \delta^2S^a$ and $\delta^2P^a$ given in (\ref{eq:SecondOrder}) 
and the relation $A_0^a=(s/c)S^a$, we can show that it actually vanishes. 
Explicitly, we have
\begin{eqnarray}
\delta^3\psi 
&=&\delta^2
\left( \half \gamma^{\mu\nu} F_{\mu\nu} - \gamma^{\mu} D_{\mu} S 
+ i  \gamma^{\mu} D_{\mu} P \gamma_5 - e [P, S] \gamma_5 \right) 
P_+(\gamma)\tilde\alpha 
\nn
&\propto&(\alpha^\dagger\sigma^i\alpha)
\Bigl(\bigl(
\bsigma\cdot\bD B^i +{e\over c}[S,\,B^i] 
\bigr)
 \otimes 
(c\sigma_1-is\sigma_3+i\sigma_2)
\Bigr)
P_+(\gamma)\tilde\alpha\,.
\end{eqnarray}
Noting the relation $c\sigma_1-is\sigma_3+i\sigma_2=i\sigma_2P_-(\gamma)$, this clearly 
vanishes, due to the orthogonality 
$P_-(\gamma)P_+(\gamma)=0$.
Thus, all the fields are zero at third order, and hence also at fourth 
order.

Now that we have obtained the finite SUSY transformation of the Julia-Zee 
dyon solution, we are able to derive the following form for  
the $U(1)$ electric field:
\begin{eqnarray}
\tilde\bE = 
\hatr^a \tilde\bE^a &=& 
-\hatr^a\bD \tilde A^a_0 
=-\hatr^a\bD (A^a_0 + {1\over2}\delta^2A_0^a) \nn
&=& -s\hatr^a\bB^a + c(1+c)\hatr^a\bD (
\alpha^\dagger\bB^a\cdot\bsigma\alpha).
\end{eqnarray} 
Then, inserting the expression for the magnetic field
\begin{equation}
B^{ia}= {\hatr^i\hatr^a\over er^2}(1-K^2) +{\delta_\perp^{ia}\over er^2}KH\ ,
\end{equation}
with the transverse Kronecker delta $\delta_\perp^{ij}\equiv\delta^{ij}-\hatr^i\hatr^j$,
we obtain
\begin{equation}
\tilde E^i = 
-{s\hatr^i\over er^2}(1-K^2) 
+ c(1+c){3\hatr^i\hatr^j-\delta^{ij}\over er^3}
(K^2H+K^2-1)(\alpha^\dagger\sigma^j\alpha)\ .
\end{equation}
The first term here is merely the Coulomb field around the 
dyon charge $Q_\e= -gs$, since we have $1/e={g/4\pi}$.
The second term is the electric dipole field induced around 
the source fermion field, $\delta^1\psi$ (\ref{eq:FirstOrder}).\footnote{%
Actually, the sources of this electric dipole field include the dipole 
moment of the gauge boson charge distribution and the EDM due to 
magnetic current, in addition to the dipole moment of the fermion 
charge distribution. 
These contributions were calculated separately 
in a previous paper\cite{Kobayashi:2006tv} 
by one of the present authors.}  
Next, noting the asymptotic behavior $H(x)/x\rightarrow1,\ K(x)\rightarrow0$ from this term, we find that the dyon fermion around the origin 
possesses the electric dipole moment
\begin{equation}
{\mbf \mu}_\e = -{4\pi c(1+c)\over e}(\alpha^\dagger{\mbf\sigma}\alpha)\,.
\end{equation}
In order to relate the operator $(\alpha^\dagger{\mbf\sigma}\alpha)$ here to the spin 
operator $\mbf J$, we must determine the normalization of our 
two-component spinor $\alpha$. For this purpose, we calculate the fermion 
number carried by our dyon fermion state $\delta^1\psi$. This is done as 
follows: 
\begin{eqnarray}
1
&=& 
\int d^3x\, (\delta^1\psi^a)^\dagger(\delta^1\psi^a) = 2c(1+c)
\int d^3x\, B^{ia}B^{aj}(\alpha^\dagger\sigma^i\sigma^j\alpha) \nn
&=& 
2c(1+c){m_{\rm dyon}\over c^2}(\alpha^\dagger\alpha)\ . 
\label{eq:NumberOp}
\end{eqnarray}
Here we have used $\int d^3x\, \bB^{a}\cdot\bB^{a} =m_{\rm dyon}/c^2$, 
which follows 
from (\ref{eq:Mass}) and (\ref{eq:SB-EBrel}). The `number operator' 
should be given by $a^\dagger a$, where 
$a^\dagger$ and $a$ are the properly normalized 
(two-component spinor) creation and annihilation 
operators, in terms of which the spin operator ${\mbf J}$ 
reads $a^\dagger({\mbf \sigma}/2)a$. Identifying the right-hand side of 
(\ref{eq:NumberOp}) with the number operator, we find
\begin{equation}
2c(1+c){m_{\rm dyon}\over c^2}\left(\alpha^\dagger{\bsigma\over2}\alpha\right) = {\mbf J}\,.
\end{equation}
In terms of this spin, the electric dipole moment of the dyon fermion is 
given by
\begin{equation}
{\mbf \mu}_\e = -{4\pi c^2\over em_{\rm dyon}}{\mbf J}
= -2c^2{g\over2m_{\rm dyon}}{\mbf J}\,,
\end{equation}
where the unit magnetic charge $4\pi/e$ is denoted by $g$. 
This result implies that the gyroelectric ratio $g_\e$ of the dyon fermion 
is $2c^2$. 

Recall that the parameter $\gamma$ is quantized as in (\ref{relation}). 
Using this, we can rewrite $c^2=\cosh^2\gamma$ in terms of the electric 
quantum number $n_\e$ and the $\theta$-parameter. Doing so, we find that 
the gyroelectric ratio 
of the dyon fermion with quantum numbers $n_\m=1$ and $n_\e$ is given by 
\begin{equation}
g_\e = 2\left(1+\Bigl({e^2\theta\over8\pi^2}+n_\e{e^2\over4\pi}\Bigr)^2\right).
\end{equation}
It is interesting that even the monopole fermion with $n_\m=1$, $n_\e=0$ 
has a gyroelectric ratio that deviates from the canonical Dirac value of 2 
in the $\theta\not=0$ case. 

It is also interesting that there are no $O(\alpha^2)$ back-reactions 
to the magnetic field, as seen from the relation $\delta^2\bA^a=0$ 
in (\ref{eq:SecondOrder}). For this reason, there is no dipole part in 
the magnetic field around the dyon, 
\begin{equation}
\bB \equiv\hatr^a\bB^a =  
{\hat{\mbf r}\over er^2}(1-K^2)\,, 
\end{equation}
and thus the asymptotic behavior is simply given by that of 
a magnetic monopole. 
For later convenience, we here write the asymptotic forms of the 
electric and magnetic fields around the monopole fermion 
(with $n_\m=1,\,n_\e=0$):
\begin{equation}
\CASE
{\bE = {e\theta/2\pi\over4\pi}{\hat{\mbf r}\over r^2}  + {4\pi/e\over m}c^2\bDdp }
{\bB = {4\pi/e\over4\pi}{\hat{\mbf r}\over r^2} + 0 \bDdp }
\qquad 
\begin{minipage}[t]{12em}  
as\ \ $r\rightarrow\infty$ \ around \\ 
\ \ \ \ \ a monopole fermion
\end{minipage}.
\label{eq:DyonAsymptField}
\end{equation}
Here $\bDdp$ is the dipole field, defined as
\begin{equation}
\bDdp \equiv 
{1\over4\pi}{3\hat{\mbf r}\,(\hat{\mbf r}{\cdot}{\mbf J})- {\mbf J}\over r^3}\  
\end{equation}
in terms of the spin operator ${\mbf J}$.

\section{$SL(2;\bZ)$ duality transformation of the field strength}

We have calculated the electric and magnetic fields around a dyon fermion 
and found the corresponding values of the electric and magnetic dipole 
moments. We next set out to obtain the dipole moments in the dual world 
from these results. 
For this purpose, we need the $SL(2;\bZ)$ 
duality transformation of the gauge field strength. 
The duality transformation of electromagnetic field has been 
known for a long time as given, e.g., in 
Refs.~\citen{Gaillard:1981rj,Gibbons:1995cv}.
Here, however, we recapitulate it in our notation in this section 
for the explicit use below.

First, recall the $SL(2;\bZ)$ duality argument of Seiberg and 
Witten.  In that argument, 
the original $SU(2)$ supersymmetric Yang-Mills system is 
described by an effective $U(1)$ supersymmetric gauge theory in the 
low energy region. Then, with the rescaling $eA_\mu\rightarrow A_\mu$, 
the action $S_\o$ of 
the $U(1)$ gauge field part of the effective theory takes 
the form
\begin{eqnarray}
-S_\o&=&{1\over32\pi}\Im \int\tau\, (F-i\tilde F)^2 = 
{1\over16\pi}\Im \int\tau\, (F^2-i\tilde F\cdot F) \nn
&=& \int\Bigl( {1\over4e^2}F^2-{\theta\over32\pi^2}\tilde F\cdot F \Bigr) ,
\label{eq:LinF}
\end{eqnarray}
where $\tau$ is defined as 
\begin{equation}
\tau\equiv{\theta\over2\pi}+i{4\pi\over e^2}\ .
\end{equation}
Note that the form given in (\ref{eq:LinF}) follows from the relations 
$\tilde{\tilde F} = - F$ and $F \cdot\tilde G = \tilde F\cdot G$ 
for the dual field strength 
$\tilde F^{\mu\nu}\equiv(1/2)\varepsilon^{\mu\nu\rho\sigma}F_{\rho\sigma}$.
Now, ignoring the definition $F_{\mu\nu}=\partial_\mu A_\nu-\partial_\nu A_\mu$, we regard 
$F_{\mu\nu}$ as an independent variable that is constrained by the Bianchi 
identity; then we have 
\begin{eqnarray}
-S&=&-S_\o+{1\over8\pi}\int V_{\D\mu}\epsilon^{\mu\nu\rho\sigma}\partial_\nu F_{\rho\sigma} 
=-S_\o+{1\over8\pi}\int\tilde F_\D \cdot F \nn
&=&-S_\o-{1\over16\pi}\Im \int(F_\D-i\tilde F_\D)(F-i\tilde F)\ ,
\end{eqnarray}
with the dual system field strength $F_{\D\mu\nu} \equiv\partial_\mu V_{\D\nu}-\partial_\nu V_{\D\mu}$.
We can now dual transform this system as follows. First, 
completing the square as 
\begin{eqnarray}
-S &=& 
{1\over32\pi}\Im \int\left[
\tau(F-i\tilde F)^2 -2(F_\D-i\tilde F_\D)(F-i\tilde F) \right] \nn
&=& 
{1\over32\pi}\Im \int\left[
\tau\left(F-i\tilde F - {1\over\tau}(F_\D-i\tilde F_\D)\right)^2 
-{1\over\tau}(F_\D-i\tilde F_\D)^2 \right]  ,
\end{eqnarray}
we perform the Gaussian integration over $F$. The same result can be 
obtained by employing the equation of motion 
\begin{equation}
\Im \left( \tau(F-i\tilde F) - (F_\D-i\tilde F_\D) \right) = 0 \ ,
\label{eq:EQM0}
\end{equation} 
which implies that the real part also vanishes so that
\begin{equation}
 \tau(F-i\tilde F) = F_\D-i\tilde F_\D \,.
\label{eq:EQM}
\end{equation} 
Thus we obtain the dual action
\begin{equation}
-S_\D =
{1\over32\pi}\Im \int\left(-{1\over\tau}\right)(F_\D-i\tilde F_\D)^2 \,.
\end{equation}
{}From (\ref{eq:EQM}), we find 
\begin{equation}
F_\D 
=\Re\left( \tau(F-i\tilde F)\right)  
=\Re\tau\,F+ \Im\tau\,\tilde F  \,.
\end{equation}
This implies that for the electromagnetic fields 
\begin{equation}
\CASE{e\bE = - (F^{01}\,F^{02},\,F^{03}),}{
e\bB=-(F^{23}\,F^{31},\,F^{12}),}
\qquad 
\CASE{e_\D\bE_\D = - (F_\D^{01}\,F_\D^{02},\,F_\D^{03}),}{
e_\D\bB_\D=-(F_\D^{23}\,F_\D^{31},\,F_\D^{12}),}
\end{equation}
we have the following mapping:
\begin{equation}
\CASE
{e_\D\bE_\D = {4\pi\over e} \bB +{e\theta\over2\pi}\bE}
{e_\D\bB_\D = -{4\pi\over e} \bE +{e\theta\over2\pi}\bB}\, .
\label{eq:1/tau}
\end{equation}
Note that from the definition $\tau_\D \equiv-1/\tau$, we have
\begin{eqnarray}
e_\D &=& \sqrt{ \Bigl({4\pi\over e} \Bigr)^2+\Bigl({e\theta\over2\pi}\Bigr)^2}
={4\pi\over e}c \ ,
 \nn
e_\D^2\theta_\D &=& -e^2\theta\ ,
\label{eq:dualEtheta}
\end{eqnarray}
with
\begin{equation}
\CASE
{
s\equiv-{e^2\theta\over8\pi^2}
}
{
c=\sqrt{1+s^2}=\sqrt{ 1 +\Bigl({e^2\theta\over8\pi^2}\Bigr)^2}
}.
\end{equation}
Here, $c$ and $s$ represent the quantities $\cosh\gamma$ and $\sinh\gamma$ 
introduced previously for the dyon solutions. 
Also note that $c = e\,e_\D/4\pi$ is a duality invariant quantity, 
as is $s$, up to its sign.

To this point, we have considered only the genuine duality transformation, 
usually called the $S$-transformation:
\begin{equation}
S:\quad \tau\ \rightarrow\ -{1\over\tau}\ .
\end{equation}
Now, let us consider the generalized 
duality transformation $SL(2;\bZ)$, which is generated by this 
$S$-transformation and the $T$-transformation 
\begin{equation}
T:\quad \tau\ \rightarrow\ \tau+1 \ ,
\end{equation}
corresponding to a shift by $2\pi$ of the $\theta$ parameter.  
The transformation is generally given by 
\begin{equation}
\tau\ \rightarrow\ \tau' = {a\tau+b\over c\tau+d}
\end{equation}
for $2\times2$ matrices $M\in SL(2;\bZ)$
\begin{equation}
M =
\mypmatrix{ a & b \\ c & d } \ ,
\quad \hbox{where}\quad 
a,b,c,d\in\bZ, \quad ad-bc=1\,.
\label{eq:SL2Z}
\end{equation}
{}Following Ref.~\citen{Gaillard:1981rj}, we identify the following 
quantity of the field strength 
\begin{equation}
\mypmatrix{
F_\D-i\tilde F_\D \equiv\tau\,(F-i\tilde F ) \\
F-i\tilde F}\ .
\label{eq:variables}
\end{equation}
as an $SL(2;\bZ)$ vector. Then, the gauge field generally transforms 
under $SL(2;\bZ)$ as 
\begin{equation}
  \mypmatrix{ F_\D-i\tilde F_\D \\ F-i\tilde F }
\ \rightarrow\  
\mypmatrix{ F'_\D-i\tilde F'_\D \\
 F'-i\tilde F' } =
M
\mypmatrix{ F_\D-i\tilde F_\D \\ F-i\tilde F }\ .
\label{eq:generalSL2Ztrf}
\end{equation}
Indeed, the relation $F_\D-i\tilde F_\D =\tau\,(F-i\tilde F )$ 
still holds after the transformation; that is, we have  
$F'_\D-i\tilde F'_\D=\tau'(F'-i\tilde F')$, which is consistent 
with the $S$ transformation discussed above,
\begin{equation}
S:\ 
  \mypmatrix{ F_\D-i\tilde F_\D \\ F-i\tilde F }
\ \rightarrow\ 
\mypmatrix{ -(F-i\tilde F) \\ F_\D-i\tilde F_\D } =
\mypmatrix{ 0 & -1 \\ 1 & 0 } 
\mypmatrix{ F_\D-i\tilde F_\D \\ F-i\tilde F }
\label{eq:Strf}
\end{equation}
where use has been made of the relation
$F_D'-i\tilde F'_D=\tau'(F'-i\tilde F')
=(-1/\tau)(F_D-i\tilde F_D)=-(F-i\tilde F)$. 
Next, note that from the definition (\ref{eq:variables}), we 
immediately obtain the 
following transformation rule under $T$: 
\begin{equation}
T:\  
  \mypmatrix{ F_\D-i\tilde F_\D \\ F-i\tilde F }
\ \rightarrow\  
\mypmatrix{ (F-i\tilde F) +  (F_\D-i\tilde F_\D) \\
 (F-i\tilde F) } =
\mypmatrix{ 1 & 1 \\ 0 & 1 } 
\mypmatrix{ F_\D-i\tilde F_\D \\ F-i\tilde F }.
\end{equation}
Transformation rule given in (\ref{eq:generalSL2Ztrf}) takes the same 
form as that of Seiberg and Witten for the complex scalar fields:
\begin{equation}
\mypmatrix{ a_\D \\ a } \ \rightarrow\ 
M \mypmatrix{ a_\D \\ a }.
\end{equation}
Here, $a$ stands for the VEV of the complex scalar field 
$a=(S^3+iP^3)$, and $a_\D$ can be identified with 
$\tau a$ similarly to Eq.~(\ref{eq:variables}).

In order to express the mass formula in terms of the 
quantum numbers $n_\e$ and $n_\m$, let us write the charges $Q_\e$ 
in (\ref{relation}) and $Q_\m=n_\m (4\pi/e)$ 
in the form
\begin{equation}
Q_\e + iQ_\m = e(n_\e +n_\m \tau) \ .
\end{equation}
Then the mass formula for the BPS states that saturate the bound 
can be written
\begin{equation}
m=a\sqrt{Q_\e^2+Q_\m^2} = ae \abs{n_\m \tau+ n_\e} = e \abs{Z}\ ,
\end{equation}
where 
\begin{equation}
Z= n_\m a_\D + n_\e a 
= (n_\m,\ n_\e)\mypmatrix{ a_\D \\ a} .
\end{equation}
In order for $Z$ to remain intact under $SL(2;\bZ)$, 
the quantum numbers should 
transform as
\begin{eqnarray}
(n_\m, \, n_\e) \ \rightarrow\ (n_\m, \, n_\e) M^{-1}\,.
\label{eq:ChargeTrs}
\end{eqnarray}

\section{$S$-duality transformation of a monopole fermion to an electron} 

Now we apply the $S$ duality transformation given in (\ref{eq:Strf}) to 
the monopole fermion state with quantum numbers $(n_\m,\, n_\e)=(1,\,0)$.
Because the quantum number is transformed to $(1,\,0) S^{-1} = (0,\,1)$ 
by the transformation appearing in (\ref{eq:ChargeTrs}), 
the monopole fermion state is transformed to the electron state with 
$(n_\m,\, n_\e)=(0,\,1)$. Therefore, we can find the asymptotic 
electromagnetic field around the electron by applying the $S$-duality 
transformation (\ref{eq:1/tau}) to the asymptotic electromagnetic field 
(\ref{eq:DyonAsymptField}) around the monopole fermion. First, we compute 
the Coulomb terms alone, deferring the dipole terms to the next step:
\begin{equation}
\CASE
{\bE_\D\Bigr|_{\rm Coulomb} = \left({4\pi\over ee_\D}{1\over e} +{e\theta\over2\pi e_\D}{e\theta/2\pi\over4\pi}\right)\times{\hat{\mbf r}\over r^2}= 
{e_\D\over4\pi}{\hat{\mbf r}\over r^2}}
{\bB_\D\Bigr|_{\rm Coulomb} = \left(-{4\pi\over ee_\D}{e\theta/2\pi\over4\pi} +{e\theta\over2\pi e_\D}{1\over e}\right)\times{\hat{\mbf r}\over r^2} = 0{\hat{\mbf r}\over r^2}} \ .
\end{equation}
Here, we have used the expression (\ref{eq:dualEtheta}) for the charge 
$e_\D$ in the dual world. This Coulomb part merely reconfirms that the 
quantum numbers change as $(1,\,0)\ \rightarrow\ (0,\,1)$ from 
the monopole fermion state to electron state under the $S$-transformation. 
Next, we calculate the dipole field part, 
which comes only from $\bE$:
\begin{equation}
\CASE
{\bE_\D\Bigr|_{\rm dipole} 
= {e\theta\over2\pi e_\D}{4\pi/e\over m}c^2\bDdp
= {e^2\theta\over8\pi^2}{e_\D\over m}\bDdp
}
{\bB_\D\Bigr|_{\rm dipole} = -{4\pi\over ee_\D}{4\pi/e\over m}c^2\bDdp = -{e_\D\over m}\bDdp}.
\label{eq:dipole}
\end{equation}
This result represents the values of the electric and magnetic dipole moments 
of the electron in the dual world. We should note that this dual 
electron world is {\em different} from the original monopole world; 
indeed, if the monopole world is a strong coupling world, i.e. $e\gg1$, 
then the 
electron world is a weak coupling world, i.e. $e_\D\ll1$, 
and the VEV of the scalar 
field is $a=cv$ in the monopole world while it is $a'=a_\D=\tau a$ in the 
electron world. Thus the 
electron mass in the dual world is given by
\begin{equation}
m_\e= e\abs{0\times a'_\D+1\times a'} = e\abs{a_\D} = m\ ,
\end{equation}
which is, as should be the case, the same as the monopole dyon mass $m$ 
in the original world. Therefore, the result for $\bB_\D\bigr|_{\rm 
dipole}$ given in (\ref{eq:dipole}) shows that the magnetic moment of 
this electron is exactly Dirac's value of $2\times(e_\D/2m)J$. It is 
interesting that the magnetic moment of the electron possesses this 
canonical value for any coupling strength of this electron world. This 
is probably due to the $N=2$ supersymmetry of the system which forbids 
higher-order radiative corrections to the magnetic moment. 

We are now ready to state the most important result of this paper. 
From the electric dipole field 
$\bE_\D\bigr|_{\rm dipole}$ given in (\ref{eq:dipole}), we find that the electron 
in this dual world carries the EDM 
\begin{equation}
{e^2\theta\over8\pi^2}{e_\D\over m}J = -{e_\D^2\theta_\D\over8\pi^2}{e_\D\over m}J \ .
\end{equation}
It is thus seen that the  EDM is equal to  
$-(e_\D^2\theta_\D/8\pi^2)$ times the Dirac magnetic moment. 

We regard this dual world as a (toy) model of our world and 
the charged, spin 1/2 fermion treated here as corresponding to electron. 
Note that the multiplet consisting of this fermion and a scalar is a 
BPS state, as is the original monopole dyon multiplet, 
and it constitutes a small supermultiplet consisting of spin 1/2 and 0, 
on which half of the supercharges vanish. 
Because $e_\D$ and $\theta_\D$ are the coupling constant and the $\theta$ 
parameter in {\em our world}, we rewrite our final result for the EDM of
the electron by dropping the subscript ``D", which indicates the dual 
world,  and using $\alpha\equiv e^2/4\pi$:
\begin{equation}
\mu_\e = -{e^2\theta\over8\pi^2}{e\over m}J
= -\alpha{\theta\over2\pi}{e\over m}J\ .
\label{eq:finalEDM} 
\end{equation}

\section{Reflection and discussion}

Considering the result (\ref{eq:finalEDM}) for the electron EDM, we 
find a strange property: This quantity is not invariant under the shift 
$\theta$, $\theta\rightarrow\theta+2\pi$. Because the system should be $2\pi$-periodic in 
$\theta$, something might be wrong. 

First, note that Witten's induced charge, $e\theta/2\pi$, for the monopole 
is obviously not periodic in $\theta$. 
However, we can always perform the $T$-transformation $\tau\rightarrow\tau+1$, 
which shifts $\theta$ by $2\pi$ and simultaneously transforms
the monopole dyon charge $(n_\m{=}1,\, n_\e)$ to 
$(n_\m{=}1, n_\e-1)$. Thus, if we shift $\theta$ by $2\pi$ and apply the 
$T$-transformation, the electric charge 
$Q_\e = n_\e e+e\theta/2\pi$ of the state is unchanged. 
Therefore, the periodicity of the system 
implies that all the states in the tower of the monopole dyons 
with $n_\m{=}1$ and $n_\e\in\bZ$ must exist in the system. For a 
given $\theta$, any one of the states in this tower of 
monopole dyons can be transformed to the electron state by applying an 
$SL(2;\bZ)$ transformation. To illustrate this, let us   
suppose that the initial state has 
the quantum numbers $n_\m{=}1$ and $n_\e$. We first perform the 
$T$-transformation $n_\e$ times, reducing the electric quantum 
number $n_e$ to zero. Then $\theta$ is changed to $\theta+2n_\e\pi$. The resulting 
state possesses the monopole's quantum numbers, i.e.  
$(n_\m, n_\e)=(1,\,0)$. Then, applying the $S$-transformation, this state 
is transformed to the electron state, with 
$(n_\m, n_\e)=(0,\,1)$. It is thus seen that applying the transformation 
$ST^{n_\e}$ to the starting state $(n_\m{=}1,\,n_\e)$, 
we can obtain the electron state, which carries the EDM
\begin{equation}
{e^2(\theta+2n_\e\pi)\over8\pi^2}{e_\D\over m}J 
= -{e_\D^2\theta_\D\over8\pi^2}{e_\D\over m}J\ ,
\end{equation}
where the coupling constant $e_\D$ and the $\theta_\D$ parameter are 
given in terms of the original parameters $e$ and $\theta$ by 
\begin{eqnarray}
e_\D &=& e \sqrt{ \Bigl({4\pi\over e^2} \Bigr)^2+\Bigl({\theta\over2\pi}+n_\e\Bigr)^2}
 = e \abs{n_\e+n_\m \tau}, \nn
\theta_\D &=& -{\theta+2n_\e\pi\over\abs{n_\e+\tau}^2} 
= -2\pi\Re\left({1\over n_\e+n_\m \tau}\right),
\end{eqnarray}
with $n_\m=1$. 
Here the result depends on the electric quantum number $n_\e$ of the 
starting state in the dyon tower. This result thus 
shows that the invariance under the shift $\theta\rightarrow\theta+2\pi$ can be realized by 
shifting $n_\e$ of the starting state in the tower. 
More importantly, the EDM of the electron state is given uniquely by the 
expression (\ref{eq:finalEDM}) if it is written in terms of the the 
coupling constant 
$e$ and the $\theta$ parameter in the electron world, in 
which the electron exists. 
Despite this invariance, we think that the starting state should be that 
with the minimum mass among those states in this dyon fermion tower 
that satisfy the condition $-1/2 < (\theta/2\pi)+n_\e \leq1/2$, 
so that the electron has the smallest possible mass. 

Despite the above considerations, the fact remains that the result given 
in (\ref{eq:finalEDM}) is not invariant under a $2\pi$ shift of $\theta$ in 
the electron world. 
However, we claim that it is not necessary 
for  the result 
to be $2\pi$-periodic in $\theta$ after the $S$-duality transformation.  
Our argument for this claim is as follows. First of all, the $\theta$-term in the $U(1)$ gauge 
theory after the $S$-duality transformation is not directly related to 
the $\theta$-term 
of a non-Abelian gauge theory, which has a topological meaning. 
The $\theta$-term in a mere $U(1)$ gauge theory is trivial, 
because $\pi_3(U(1))=0$, and hence no periodicity in $\theta$ is necessary. 
In the present electron world considered presently, 
the $\theta$-term, which is proportional to $\theta{\mbf E}\cdot {\mbf B}$,  
has the following meaning that it causes magnetically charged objects to 
possess the Witten charge. Indeed, this system has a BPS 
monopole multiplet with $n_\m{=}\pm1, n_\e{=}0$, which is obtained by 
the $S$-duality transformation of the 
$W$-boson multiplet. Thus measuring the electric charge $Q_\e=e\theta/2\pi$ 
of this monopole multiplet, we should be able to determine the value of 
$\theta$.\footnote{%
As we claim that there is no necessity for the periodicity of $\mu_\e$ 
in $\theta$, we also believe that there exists no $T$-transformation tower 
states of the monopole, with quantum numbers $n_\m{=}\pm1$, $n_\e\in \bZ$. 
We thus believe that the electric charge $Q_\e$ of the monopole multiplet 
is given solely by the Witten effect.}

We called the fermion dual to the monopole fermion, `electron'. 
However, if we seriously want to identify this object with the 
real electron, there are many problems. First of all, the present 
computation heavily 
depends on the existence of $N=2$ supersymmetry. We do not know how 
much our computation can be generalized to the systems with $N=1$ 
or no supersymmetry. Moreover, it is completely unclear at the moment 
how the usual Standard Model with quarks and leptons can be 
reconstructed or embedded in such a dual world in which our real 
electron is identified with an object dual to the monopole fermion in a 
certain non-Abelian gauge theory.

\section*{Acknowledgements}
The authors would like to thank Tohru Eguchi, Ken-Ichi Izawa, 
Ken-Ichi Konishi, Hiroshi Kunitomo, Ken-Ichi Shizuya, Haruhiko Terao and
Seiji Terashima for valuable discussions. 
M.K.\ is partially supported by a Grant-in-Aid for Scientific Research on 
Priority Areas (No.\ 13135225) from the Ministry of Education, Culture, 
Sports, Science and Technology (MEXT). 
T.K.\ and T.T.\ are partially supported by a Grant-in-Aid for Scientific 
Research (B) (No.\ 16340071) and by Grants-in-Aid for JSPS Fellows, 
respectively, from the Japan Society for the Promotion of Science. 
They are also supported by a Grant-in-Aid for the 21st Century COE 
``Center for Diversity and Universality in Physics".

\end{document}